\documentclass[%
 amsmath,amssymb,
 aps,
 twocolumn
]{revtex4-2}

\usepackage{graphicx}
\usepackage{dcolumn}
\usepackage{bm}
\usepackage[colorlinks=true,citecolor=blue]{hyperref}
\usepackage{url}

\begin{document}

\title{Study of the mass of pseudoscalar glueball with a deep neural network}

\author{Lin Gao}
\email{silvester\_gao@qq.com}

\affiliation{American Association for the Advancement of Science, Washington, DC 20005, USA}

\date{\today}

\begin{abstract}

A deep neural network (DNN) is utilized to study the mass of the pseudoscalar glueball in lattice QCD based on Monte Carlo simulations. The DNN is constructed to extract the mass from the negative part of the topological charge density correlation function. The resulting mass estimates are compared with those obtained from conventional least-squares fitting. The DNN gives a pseudoscalar glueball mass of $2558(89)\,\mathrm{MeV}$, while the conventional fit gives $2612(112)\,\mathrm{MeV}$. The results suggest that the DNN provides a stable and complementary approach to conventional mass extraction from lattice correlation functions.

\end{abstract}

\maketitle

\section{Introduction}
Glueballs, the bound states of gluons predicted by Gell-Mann in 1962 \cite{GellMann1962}, serve as a crucial element in Quantum Chromodynamics (QCD), where gluons act as the mediators of strong interactions, described by the SU(3) gauge fields \( A_\mu(x)_{cd} \). Employing Euclidean spacetime, with Lorentz indices \(\mu\) ranging from \(\mu = 1, 2, 3, 4\), and color indices \( c \) and \( d \) spanning \( c, d = 1, 2, 3 \), the gauge field \( A_\mu(x) \) can be expressed as\cite{Gattringer2010}
\begin{equation}
A_\mu(x) = \sum_{i=1}^{8} A_\mu^i(x) T_i,
\end{equation}
where \( T_i \) are the generators of SU(3), and the resulting \( A_\mu(x) \) are Hermitian and traceless matrices. In contrast to the single type of photon, mediating electromagnetic interactions in Quantum Electrodynamics, gluons exhibit eight types, each carrying color charge, thus engendering self-interactions and complicating the analysis of strong interactions.

Theoretically, several gluons can interact to form relatively stable bound states termed glueballs. These glueballs can be classified by their quantum numbers, including spin \( J \), parity \( P \), and charge conjugation \( C \). For instance, the quantum numbers of a pseudoscalar glueball are denoted as \( J^{PC} = 0^{-+} \).

Understanding glueballs is crucial for deepening our comprehension of non-perturbative QCD effects and the confinement phenomenon. This paper aims to study the mass of pseudoscalar glueball. To achieve this, it is necessary to introduce the topological charge density\cite{Gattringer2010}
\begin{equation}
q\left(x\right)=\frac{1}{32\pi^2}\varepsilon_{\mu \nu\rho\sigma}{\rm{Re}}{{\rm{Tr}}{\left[F_{\mu\nu}\left(x\right)F_{\rho\sigma}\left(x\right)\right]}},
\end{equation}
where  \( F_{\mu\nu}(x) \) is the field strength tensor and the correlation function of  \( q(x) \) is as follows
\begin{equation}
C_{q}(r)=\langle q(x)q(y)\rangle,  \quad  r=|x-y|.
\end{equation}

In lattice QCD, the pseudoscalar glueball mass can be extracted from the long-distance behavior of the two-point correlation function constructed from the topological charge density. The negative tail of the topological charge
density correlation function can be written as\cite{Chowdhury_2015}
\begin{equation}
\widetilde{C}_q(r)
=
A\frac{m}{4\pi^2r}K_1(mr),
\end{equation}
where \( m \) is the pseudoscalar glueball mass, \( A \) is a normalization factor,$\widetilde{C}_q(r)<0$ and $K_1$ is the modified Bessel function whose asymptotic form at large z is given by
\begin{equation}
K_1(z) \sim K_{1A}(z) = \sqrt{\frac{\pi}{2z}} \mathrm{e}^{-z} \left(1 + \frac{3}{8z}\right).
\label{eq:K1A_definition}
\end{equation}
This asymptotic behavior shows that the $K_1$ form contains the
exponential decay together with additional power-law dependence on the
distance.

In this work, the topological charge density is evaluated on the lattice using the
clover discretization of the field strength tensor $F_{\mu\nu}\left(x\right)$ and it can be written as\cite{Gattringer2010}
\begin{footnotesize}
\begin{equation}
\begin{array}{*{20}{c}}
F_{\mu\nu}^{clov}(n)=-\frac{i}{8a^2}\left[\left(C_{\mu\nu}\left(n\right)-C_{\mu\nu}^\dag(n)\right)-\frac{{\rm{Tr}}\left(C_{\mu\nu}\left(n\right)-C_{\mu\nu}^\dag(n)\right)}{3}\right],
\end{array}
\end{equation}
\end{footnotesize}
where the clover is given as 
\begin{equation}
C_{\mu\nu}(n)=U_{\mu,\nu}(n)+U_{\nu,-\mu}(n)+U_{-\mu,-\nu}(n)+U_{-\nu,\mu}(n).
\end{equation}
The $U_{\mu v}\left(n\right)$ is the plaquette.
The resulting lattice correlation function $C_q(r_n)$ and its negative
tail $\widetilde{C}_q(r_n)$ are then obtained, where $r_n$ denotes the
discrete lattice separation.

On the other hand, machine learning (ML) is advancing the study of QCD and has been applied to a wide range of problems in particle physics \cite{Giarnetti2026}.
For instance, modified generative adversarial network approaches, including a DCGAN and a Wasserstein GAN, have been utilized to study topological quantities in lattice QCD \cite{gao2024dcgan,gao2024}. 
The extraction of the pseudoscalar glueball mass from the $K_1$
correlation function is a nonlinear fitting problem. Conventional
least-squares(LS) fitting determines the parameters by directly minimizing
the deviations between the model and the data. Its performance can
become sensitive to statistical fluctuations, particularly when the
available correlation-function data are limited or noisy. Machine
learning provides an alternative approach by learning the mapping
between the correlation function and its underlying mass parameter.
In this work, a Deep Neural Network (DNN) is therefore employed to
extract the pseudoscalar glueball mass from the correlation-function
data. Noise is intentionally introduced into the training data to
improve the robustness of the trained model. The results obtained by
the DNN are compared with those from the conventional
$K_1$ least-squares fits, allowing the effectiveness of the machine
learning approach to be systematically investigated.

\section{Data preparation}
In this section, the aim is to obtain the correlation function data of the topological charge density \( q(x) \). These data will be used for both traditional LS method and ML approach to determine the mass of the pseudoscalar glueball.

Firstly, the pseudo heat bath algorithm with periodic boundary conditions is employed to simulate lattice QCD configurations\cite{Gattringer2010}. Subsequently, the Wilson flow is applied to smear these configurations\cite{Luscher2010}. The action employed here is the Wilson gauge action\cite{Gattringer2010}
\begin{equation}
{S_G}\left[ U \right] = \frac{\beta }{3}\mathop \sum \limits_{n \in {\rm{\Lambda }}} \mathop \sum \limits_{\mu  < \nu} {\mathop{\rm Re}\nolimits} {\rm{Tr}}\left[ {1 - {U_{\mu \nu}}\left( n \right)} \right],
\end{equation}
where $\beta$ is the inverse coupling and $U_{\mu \nu}\left(n\right)$ is the plaquette. Configurations are generated using the software Chroma\cite{Edwards_2005} on a personal workstation. Due to the high computational cost of computing the sum of the staples, the updating steps are repeated 10 times for the visited link variable. After the system reaches equilibrium, a total of 1000 configurations are sampled at intervals of 200 sweeps each. The integrated autocorrelation time for topological charge calculation is 0.416, indicating that each of data can be considered independent\cite{gao2024}.

Furthermore, the static QCD potential is utilized to set the scale and can be parameterized by\cite{Gattringer2010}
\begin{equation}
V(r)=A+\frac{B}{r}+\sigma r.
\end{equation}
The Sommer parameter \( r_0 \) is defined as
\begin{equation}
 r_0^2 \frac{dV(r)}{dr}\bigg|_{r=r_0} = 1.65, 
\end{equation}
and \( r_0 = 0.49 \) fm is used\cite{Sommer2014scale15}. The scale is shown in Tab.~\ref{static_potential_scale}.
\begin{table}[htb]
\caption{\label{static_potential_scale}The scale set through the static QCD potential.}
\begin{ruledtabular}
\begin{tabular}{cccccc}
\textrm{$Volume$}&
\textrm{$\beta$}&
\textrm{$a[fm]$}&
\textrm{$N_{cnfg}$}&
\textrm{${r_0}/a$}&
\textrm{$L[fm]$}\\
\colrule
$24\times{12}^3$ & 6.0&0.093(3)&1000&5.30(15)&1.11(3) \\
\end{tabular}
\end{ruledtabular}
\end{table}

The lattice QCD configurations are further processed using Wilson flow \cite{Luscher2010}. The step size of the Wilson flow is set to \( \varepsilon_f = 0.01 \). If the Wilson flow time \(t_f\) is too large, it can lead to the elimination of negative part \( \widetilde{C}_q(r_n) \) of the correlation function \( C_q(r_n) \). Therefore, an appropriate Wilson flow time \( t_f \) needs to be determined. In this paper, I adopt \( \sqrt{8t_f} = 0.15 \) fm \cite{Chowdhury_2015}. Subsequently, the smeared configurations are used to compute \( C_q(r_n) \).

\section{ \(K_1\)-Based Conventional Least-Squares Fit}

\subsection{Asymptotic Parametrization}

For sufficiently large $mr$, the modified Bessel function appearing in the
correlation function can be approximated by its asymptotic form,
\begin{equation}
K_1(mr) \sim K_{1A}(mr),
\end{equation}
where $K_{1A}$ is defined in Eq.~\eqref{eq:K1A_definition}.
Accordingly, the fitting function used in the present analysis is
\begin{equation}
f_1(r;A,m)
=
A\frac{m}{4\pi^2r}K_{1A}(mr).
\label{eq:asymptotic_fit}
\end{equation}
Here, $A$ is the amplitude and $m$ is the mass parameter to be
determined from the fit.

\subsection{Least-Squares Fitting}
A total of 1000 configurations were simulated, from which the correlation
function $C_q(r_n)$ was calculated. A least-squares fit to the resulting
correlation function was then performed, yielding the pseudoscalar glueball
mass m=2612(112)MeV. Further details will be discussed later in the article.

\section{Machine learning model}

\subsection{Structure}

DNN can assist in studying the mass of pseudoscalar glueball in lattice QCD. Tests have shown that when the structure of the DNN is too simple, the mass of the pseudoscalar glueball obtained from different amounts of data is inaccurate and unstable. Therefore, a more complex DNN was constructed. The structure of this DNN is illustrated in Fig.~\ref{DNN_structrue_ref}. This neural network employs 8 fully connected layers, each activated by a LeakyReLU function, except for the final layer, which uses a Sigmoid function. More specific parameters are detailed in the Tab.~\ref{DNN_structure_tab}.
\begin{figure*}[htb]
    \centering
    \includegraphics[width=0.8\textwidth]{DNN_structrue}
    \caption{The structure of the DNN used to obtain the mass of the pseudoscalar glueball.}
    \label{DNN_structrue_ref}
\end{figure*}

In the application phase, the inputs and outputs of the DNN are explained as follows. The input is a sequence \(\left(c_1, c_2, \cdots, c_N\right)\) composed of the negative values \(\widetilde{C}_q\left(r_n\right)\) of the correlation function  calculated from lattice QCD at different positions. The value of \(N\) varies for different lattice sizes, and in this paper, \(N\) is 30. The output consists of two parameters, \(P_1\) and \(P_2\), both within the range (0, 1). The Sigmoid function in the last layer ensures that the output is within this range. These output parameters, \(P_1\) and \(P_2\), are proportional to the parameters \(A\) and \(m\) in \(\widetilde{C}_q\left(r_n\right)\). Since the range of the DNN's output is (0, 1), the mass \(m\) of the pseudoscalar glueball needs to be scaled to fall within this range, corresponding to \(P_2\). In the present calculation, the training range of the dimensionless mass parameter is restricted to $0<aM<2$, so that the corresponding normalized output aM/2 lies in $(0,1)$. Similarly, the parameter \(A\) in \(\widetilde{C}_q\left(r\right)\) is scaled accordingly. Therefore, the output parameter \(P_1\) is proportional to the parameter \(A\), and \(P_2\) corresponds to \(aM/2\). By obtaining the value of \(aM/2\) from the DNN, the value of \(M\) can be further determined. The detailed procedure on how the DNN is trained to ensure that \(P_1\) and \(P_2\) are proportional to the parameters \(A\) and \(m\) given the input \(\left(c_1, c_2, \cdots, c_N\right)\) is elaborated in the training phase of the DNN.

The explanation for the parameter number in Tab.~\ref{DNN_structure_tab} is as follows. Taking Linear-1 as an example, the input to Linear-1 is a tensor with the shape \([-1, 30]\), where \(-1\) is a placeholder corresponding to the batch size. Linear-1 has 6400 neurons, and each neuron has a bias. Therefore, the Parameter number is calculated as \(30 \times 6400 + 6400 = 198,400\).

\begin{table}[htb]
\caption{\label{DNN_structure_tab}The structure of the DNN.}
\begin{ruledtabular}
\begin{tabular}{llr}
\textrm{Layer (type) }&
\textrm{Output Shape }&
\textrm{Parameter number }\\
\colrule
Linear-1&[-1, 6400]&198,400\\
LeakyReLU-2&[-1, 6400]&0\\
Linear-3&[-1, 100]&640,100\\
LeakyReLU-4&[-1, 100]&0\\
Linear-5&[-1, 4]&404\\
LeakyReLU-6&[-1, 4]&0\\
Linear-7&[-1, 100]&500\\
LeakyReLU-8&[-1, 100]&0\\
Linear-9&[-1, 4]&404\\
LeakyReLU-10&[-1, 4]&0\\
Linear-11&[-1, 100]&500\\
LeakyReLU-12&[-1, 100]&0\\
Linear-13&[-1, 6400]&646,400\\
LeakyReLU-14&[-1, 6400]&0\\
Linear-15&[-1, 2]&12,802\\
Sigmoid-16&[-1, 2]&0\\
\end{tabular}
\end{ruledtabular}
\end{table}

\subsection{Training}
\begin{figure}[htb]
\includegraphics[width=0.4\textwidth]{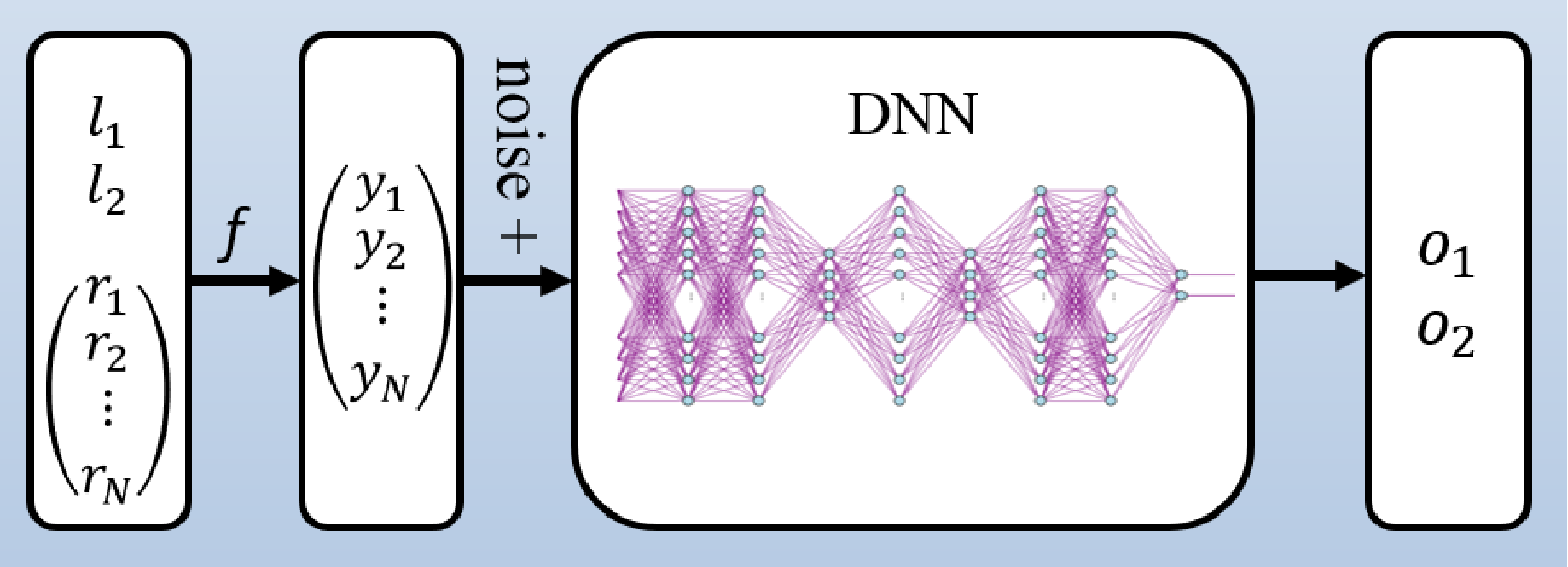}
\caption{\label{DNN_training_ref}DNN training process.}
\end{figure}

To enable the neural network to accurately determine the mass of the pseudoscalar glueball, the following method was used to train the DNN. The training process is illustrated in Fig.~\ref{DNN_training_ref}. During the training phase, the following function was introduced
\begin{equation}
f(r_n, l_1, l_2) = -\frac{50 l_1 l_2}{\pi^2 a r_n} K_{1A}\left(\frac{2 l_2 r_n}{a}\right),
\end{equation}
where \(K_{1A}(z)\) is the asymptotic form of the modified Bessel function at large positive \(z\). Specifically, when \(l_1 = -A/100\) and \(l_2 = am/2\), it follows that \(\widetilde{C}_q(r_n) = f(r_n, -A/100, am/2)\).

For the preparation of the training data, two values were randomly selected from a uniform distribution between 0 and 1 as \(l_1\) and \(l_2\). These values, along with \(r_n\), were then used to compute \(y_n = f(r_n, l_1, l_2)\), resulting in an array \(\left(y_1, y_2, \cdots, y_N\right)\), to which noise was added. A total of 10,000 such processed arrays constituted the training dataset. From this dataset, arrays with batch size = 100 were selected and input into the DNN, which was trained to produce outputs \(o_1\) and \(o_2\) matching \(l_1\) and \(l_2\). Essentially, \(l_1\) and \(l_2\) serve as the labels for the arrays \(\left(y_1, y_2, \cdots, y_N\right)\), and the goal of the DNN is to output the correct labels given the input arrays.

The \(\widetilde{C}_q(r_n)\) is assumed to be approximately described by the
same functional form \(f(r_n, l_1, l_2)\) within the fitting range. Under this assumption,
the trained DNN can be applied to the lattice correlation-function data
to infer the \(l_1\) and \(l_2\), and consequently determine the parameters \(A\) and the mass \(m\), when given the \(\widetilde{C}_q(r_n)\) data computed from lattice QCD.

The optimizer used is Adam, proposed by Kingma and Ba \cite{kingma2014adam}, which combines the advantages of adaptive learning rates and momentum. Adam is chosen as the appropriate optimizer in this study due to its low memory requirements, automatic adjustment of learning rates, and ability to keep the step sizes within a general range.The loss function used in this paper is Binary Cross Entropy. Additionally, some programs are implemented using PyTorch \cite{PASZKE2019PyTorch}.

\section{Results}
\begin{figure}[htb]
\includegraphics[width=0.5\textwidth]{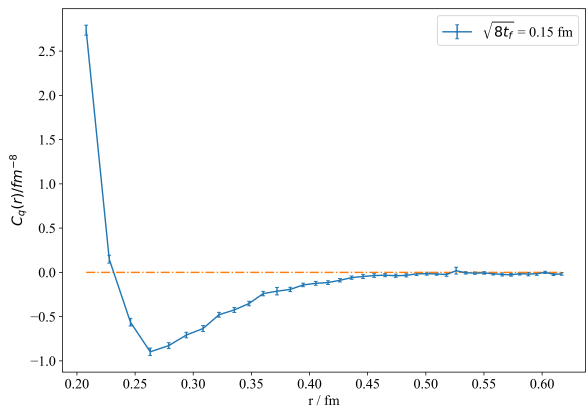}
\caption{\label{correlation_curve_ref}The curve of the correlation function \(C_q\left(r\right)\) with \(\sqrt{8t_f} = 0.15 \, \mathrm{fm}\).}
\end{figure}

First, the correlation function of the topological charge density calculated by the Monte Carlo method with Wilson flow is presented. As shown in Fig.~\ref{correlation_curve_ref}, it can be observed that in the region where \( r \) is relatively large, the value of the correlation function increases with \( r \) and gradually approaches 0 from negative values, aligning with the characteristics of the negative part \( \widetilde{C}_q(r) \) in \( C_q(r) \).

 For the LS method, I simulated a total of 1000 configurations and calculated the correlation function $C_q\left(r_n\right)$ from these configurations. Then the mass of the pseudoscalar glueball m=2612(112)MeV was fitted from 1000 ${\widetilde{C}}_q\left(r_n\right)$ data and details of the calculations are shown in Tab.~\ref{LS_and_ML}, when 25 CPU cores were used in calculations. 

For the DNN, noise was added to the training data during
the training process. The resulting model was then applied to
$\widetilde{C}_q(r_n)$ with different amounts of data to determine the
mass of the pseudoscalar glueball.Using 1000
$\widetilde{C}_q(r_n)$ data points, the model trained with noise-augmented
data yielded a mass of $m = 2558(89)\,\mathrm{MeV}$, which is consistent
with the results obtained by other researchers\cite{Chowdhury_2015}.

In terms of data details, the negative part of the correlation function,
\(\widetilde{C}_q(r_n)\), is used to construct  sequences \((c_1, c_2, \cdots, c_N)\) at different distances \(r_n\).
 Traditional LS method and ML method were then used to determine the mass of the pseudoscalar glueball, with errors handled using the jackknife method. The results are shown in Tab.~\ref{LS_and_ML}.

\begin{figure}[htb]
\includegraphics[width=0.5\textwidth]{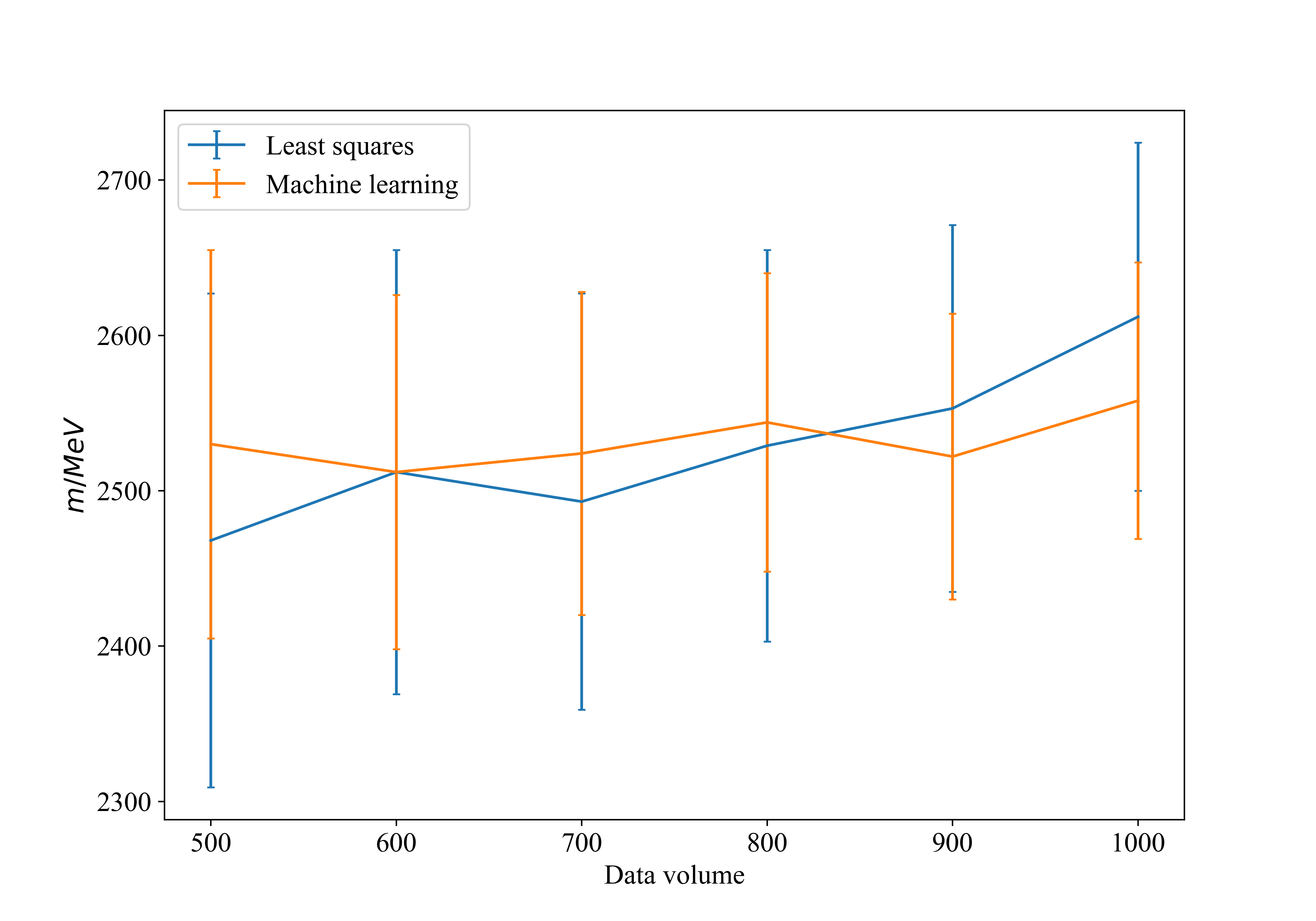}
\caption{\label{glueball_ML_LS_ref}The mass of the pseudoscalar glueball obtained using the traditional LS method and the ML approach under different data volumes}
\end{figure}

The masses obtained from the conventional LS method and the DNN are summarized in Table~\ref{LS_and_ML} and Fig.~\ref{glueball_ML_LS_ref}.
The data error gradually decreases as the volume of data increases for both LS and DNN. In cases with the same amount of data, ML yields smaller errors in determining the mass of the pseudoscalar glueball compared to the LS. Furthermore, the fitting results obtained by ML are more stable across different data volumes. 
\begin{table}[htb]
\caption{\label{LS_and_ML}The mass $m$ of the pseudoscalar glueball obtained using the traditional LS method and the ML method.}
\begin{ruledtabular}
\begin{tabular}{ccc}
\textrm{Data volume}&
\textrm{$m(ML)/MeV$}&
\textrm{$m (LS)/MeV$}\\
\colrule
500 & 2530(125) & 2468(159)  \\
600 & 2512(114) & 2512(143)  \\
700 & 2524(104) & 2493(134)  \\
800 & 2544(96) & 2529(126)  \\
900 & 2522(92) & 2553(118)  \\
1000 & 2558(89) & 2612(112)  \\
\end{tabular}
\end{ruledtabular}
\end{table}

One possible explanation for these results is as follows. Fully connected layers in this DNN, by connecting all input units, can learn the global features of the input data. Despite the presence of noise in the data, the fully connected layers can still identify useful patterns. During training, the network adjusts its weights to fit the effective patterns in the data while ignoring the noisy features as much as possible. The stacking of multiple fully connected layers enhances the network's nonlinear representation capability, allowing it to differentiate useful features from noise to some extent. Through this multi-layer structure, the network can extract higher-level features layer by layer, which helps to filter out the impact of noise at higher levels. Therefore, DNNs can suppress the influence of noise and extract effective features from the data to a certain degree. Related research also shows that DNNs can filter out noise in the data\cite{kolbaek2018}. In contrast, the LS method determines model parameters by minimizing the sum of the squared errors between the original data and the predicted data, which makes it more susceptible to the impact of noise when the data volume is small, leading to instability in the results.

\section{Conclusion}
This work investigates the pseudoscalar glueball mass from the long-distance behavior of the topological charge density correlation function in lattice QCD. Two approaches are considered: a conventional least-squares fit based on the asymptotic form of the modified Bessel function $K_1$, and a fully connected deep neural network (DNN) method.

A conventional least-squares fit using the asymptotic $K_1$ parametrization gives a pseudoscalar glueball mass of $M_{0^{-+}}=2612(112),\mathrm{MeV}$. This provides a direct determination of the mass from the long-distance behavior of the topological charge density correlation function while accounting for the radial dependence of the four-dimensional Euclidean correlator.

In addition, a fully connected DNN is employed to extract the pseudoscalar glueball mass from the correlation-function data. For 1000 correlation-function data points, the DNN gives $M_{0^{-+}}=2558(89),\mathrm{MeV}$. The DNN result is consistent with the conventional $K_1$-based fit within the statistical uncertainties, while yielding a smaller statistical uncertainty for the data volume considered in this study.

Overall, the two approaches considered in this work give consistent estimates of the pseudoscalar glueball mass. The agreement between the conventional $K_1$-based least-squares fit and the DNN approach demonstrates the potential of combining conventional correlation-function analysis with machine-learning methods for the extraction of physical observables in lattice QCD.

\textbf{Acknowledgments.}  This research utilized Chroma for simulating the lattice gauge field configurations. I extend my gratitude to the contributors of Chroma for their valuable contributions.

\bibliographystyle{apsrev4-2}
\bibliography{apssamp}

\end{document}